\RequirePackage{luatex85}
\documentclass[english,aps,prl,twocolumn,showpacs,superscriptaddress,groupedaddress,fixfloats]{revtex4-1}

    \usepackage{parskip}
    \usepackage{physics}
    \usepackage{amsmath}
    \usepackage{amssymb}
    \usepackage{xcolor}
    \usepackage[colorlinks,breaklinks=true]{hyperref}
    \usepackage{array}
    \usepackage{longtable}
    \usepackage{multirow}
    \usepackage{comment}
    \usepackage{graphicx}
    \usepackage{amsfonts}
    \usepackage{bm}
    \usepackage{slashed}
    \usepackage{dsfont}
	\usepackage{mathtools}
	\usepackage[compat=1.1.0]{tikz-feynman}
	\usepackage{makecell}
    \usepackage{mathrsfs}
    \usepackage{xparse}
	\usepackage{enumerate}
	\usepackage{mathtools}

    \usepackage[T1]{fontenc}
    \usepackage[caption=false]{subfig}
	
    \usepackage{esint}
    \usepackage{bbm}
    \usepackage{babel}
    \usepackage{footmisc}

	\newcommand{\nn}{\nonumber}
	
	\newcommand{\beq}{\begin{equation}}
	\newcommand{\eeq}{\end{equation}}
	\newcommand{\bqa}{\begin{eqnarray}}
	\newcommand{\eqa}{\end{eqnarray}}
	\newcommand {\bseq}{\begin{subequations}}
	\newcommand {\eseq}{\end{subequations}}
	
	\newcommand*{\rom}[1]{\uppercase\expandafter{\romannumeral #1\relax}}
        
        \def\XXint#1#2#3{{\setbox0=\hbox{$#1{#2#3}{\int}$ }
        \vcenter{\hbox{$#2#3$ }}\kern-.6\wd0}}

    \usepackage{titlesec}
    \titleformat{\paragraph}[runin]{\itshape}{}{}{}[]
    \titlespacing*{\paragraph}{0pt}{*}{*}

\makeatletter
\newcommand{\pushright}[1]{\ifmeasuring@#1\else\omit\hfill$\displaystyle#1$\fi\ignorespaces}
\newcommand{\pushleft}[1]{\ifmeasuring@#1\else\omit$\displaystyle#1$\hfill\fi\ignorespaces}
\makeatother

\usepackage{shellesc}
\usetikzlibrary{external}

\newcommand{\dotsize}{2.5mm}

\tikzfeynmanset{
	HQET/.style={
		/tikz/draw=none,
		/tikz/decoration={name=none},
		/tikz/postaction={
			/tikz/draw,
			/tikz/double distance=1.5pt,
		},
	},
	qed/.style={
		/tikz/postaction={
			/tikz/decoration={
				markings,
				mark=at position 0.5 with {
					\node[
					transform shape,
					xshift=-0.5mm,
					fill,
					inner sep=1.3pt,
					draw=none,
					isosceles triangle
					] { };
				},
			},
			/tikz/decorate=true,
		},
	},
	with arrow/.style={
		/tikz/decoration={
				markings,
				mark=at position #1 with {
					\node[
					transform shape,
					xshift=-0.1mm,
					fill,
					dart tail angle=120,
					inner sep=1.1pt,
					draw=none,
					dart
					] { };
				},
			},
		/tikz/postaction={
			/tikz/decorate=true,
		},
	},
	empty square dot/.style={
		/tikzfeynman/square dot,
		/tikz/minimum size=0.7mm,
		/tikz/white,
	},
	my blob/.style={
		/tikzfeynman/blob,
		/tikz/minimum size=30pt,
	},
	momentum/arrow distance=2mm,
}

\begin{document}


\title{Near-the-origin divergence of Dirac wave functions of hydrogen
and operator product expansion}

\author{Yingsheng Huang}
\email{huangys@ihep.ac.cn}
\affiliation{Institute of High Energy Physics, Chinese Academy of
	Sciences, Beijing 100049, China\vspace{0.2 cm}}
\affiliation{School of Physics, University of Chinese Academy of Sciences,
	Beijing 100049, China\vspace{0.2 cm}}

\author{Yu Jia}
\email{jiay@ihep.ac.cn}
\affiliation{Institute of High Energy Physics, Chinese Academy of
	Sciences, Beijing 100049, China\vspace{0.2 cm}}
\affiliation{School of Physics, University of Chinese Academy of Sciences,
	Beijing 100049, China\vspace{0.2 cm}}

\author{Rui Yu}
\email{yurui@ihep.ac.cn}
\affiliation{Institute of High Energy Physics, Chinese Academy of Sciences, Beijing 100049, China\vspace{0.2 cm}}
\affiliation{School of Physics, University of Chinese Academy of Sciences,
	Beijing 100049, China\vspace{0.2 cm}}

\date{\today}

\begin{abstract}
There is a long-standing puzzle concerning the Coulomb solutions of the Dirac equation, {\it i.e.},
what is the physics governing the weakly divergent near-the-origin behavior of the
Dirac wave functions of the $nS_{1/2}$ hydrogen?
As a sequel of our preceding work that aim to demystifying the universal near-the-origin behavior
of the atomic Schr\"{o}dinger and Klein-Gordon wave functions~\cite{Huang:2018yyf,Huang:2018ils},
the goal of this work is to demonstrate that, within the nonrelativistic effective field theory (NREFT)
tailored for Coulombic atoms, the universal logarithmic divergence of
the Dirac wave functions can be accounted by the {\it perturbatively calculable}
Wilson coefficient emerging from the operator product expansion (OPE) of the electron and the nucleus fields.
The cause is due to the relativistic kinetic correction and Darwin (zitterbewegung) term in the NREFT.
With the aid of renormalization group equation, one can resum the leading logarithms to all orders in $Z\alpha$
and recover the $r^{-Z^2\alpha^2/2}$ anomalous scaling behavior
exhibited by the Dirac wave function for the $nS_{1/2}$ hydrogen.
It appears somewhat counterintuitive that these universal logarithmic divergences
can not be accounted by the OPE set up in the relativistic QED. We are thereby enforced to conclude that
the Dirac wave function must cease to be meaningful when $r$ is shorter
than the electron's Compton wavelength.
\end{abstract}

\maketitle
\paragraph{\color{blue}Introduction.}
Dirac equation~\cite{Dirac:1928hu} provides a successful
relativistic quantum mechanical framework for the spin-${1\over 2}$ electron,
often ranked among the top celebrated equations in all time of physics.
The Coulomb solution of the Dirac equation, {\it e.g.}, a relativistic wave equation for an
electron immersed in an electrostatic Coulomb field,
has predicted the correct fine structure of the hydrogen spectrum
as early as in 1928~\cite{Darwin1928,Gordon1928}.
It is difficult to underestimate the importance of this historic achievement,
which is often viewed as one of greatest triumph of (relativistic) quantum mechanics.

As is widely known, the Coulomb solutions of the Dirac equation scales as $r^{\nu-1}$ near the origin,
with $\nu^2=\left(j+{1\over 2}\right)^2-Z^2\alpha^2$~\cite{Bethe1977,landau:1982}.
A curious conundrum then arises. Whereas the Schr\"{o}dinger wave functions of hydrogen are always
regular at the origin, the corresponding $n S_{1/2}$ Dirac wave functions become weakly (logarithmically)
divergent at the origin~\cite{Bethe1977}. How this pattern can arise?
What is the physics behind this near-the-origin divergence of Dirac wave function?
It is often heard that since the spatial integral of $\Psi^\dagger\Psi$ still remains finite,
this divergence at the origin appear not to violate probabilistic interpretation of wave function.
Moreover, this singular behavior of Dirac wave function brings pronounced effect only in extremely small $r$,
actually many orders-of-magnitude smaller than the charge radius of the nucleus. Therefore,
this divergence seems not to cause any practical disaster.
However, from the theoretical angle, it is of great curiosity to unravel the myth
underlying this weak divergent behavior near the origin.

Surprisingly, the puzzle related to the near-the-origin divergence has never been satisfactorily
addressed in the past 90 years. There exist some sporadic work that attempted to understand this myth in
the context of relativistic Coulomb bound state formalism~(for an incomplete list, see \cite{Durand:1983dc,Castorina:1984rd,Malenfant:1987tm,Malenfant:1988zw,Chen2008}).
Unfortunately all of them tend to regard this divergence as a nuisance, and have not
realized its perturbative nature.

The goal of this Letter is to demonstrate that this long-standing puzzle
in relativistic wave equation can be resolved by appealing to some modern field-theoretical tools
in particle physics.
We show that, within the framework of nonrelativistic effective field theory (NREFT),
the universal logarithmic divergences of the hydrogen Dirac wave functions can be accounted by
the {\it perturbatively calculable} Wilson coefficient affiliated with the operator product expansion (OPE)~\cite{Wilson:1969zs}.
An important conceptual clarification is that the Dirac wave function probed at a length scale shorter than the
electron Compton wavelength may become meaningless.

This Letter, which aims at deciphering the universal short-distance behavior of
Dirac wave function, together with our two recent work~\cite{Huang:2018yyf,Huang:2018ils},
which also apply NREFT and OPE to account for the universal electron-nucleus coalescence behavior
of the atomic Schr\"{o}dinger and Klein-Gordon wave functions, formed a trilogy.
The current work in many aspects resembles the Klein-Gordon case~\cite{Huang:2018ils},
to which we refer the interested readers for greater technical details.

\paragraph{\color{blue} Divergence of Dirac wave functions for $nS_{1/2}$ hydrogen.}
The Dirac equation in a Coulomb field of a fixed nucleus with charge $Ze$
reads
\begin{align}
\left(-i\hbar c\bm{\alpha}\cdot \bm{\nabla} +\beta m c^2- {Z e^2\over 4\pi r} \right) \Psi=E\Psi,
\label{Dirac:equation:Coulomb:potential}
\end{align}
where $\bm{\alpha},\beta$ are the Dirac matrices, $m$ is the electron mass.
In the nonrelativistic limit ($Z\alpha\ll 1$),
the energy spectrum for the hydrogen-like atoms are approximately
\begin{align}	
E_{nj} = m c^2 \left[1-{Z^2\alpha^2\over 2n^2}- {Z^4\alpha^4\over 2n^4}\left({n\over j+
{1\over 2}}-\frac{3}{4}\right)+\cdots\right],
\label{energy:level:hydrogen}
\end{align}
where $\alpha\equiv {e^2\over 4\pi\hbar c}$ signifies the fine structure constant.

From now on we will concentrate on the Dirac wave functions
for $j=\frac{1}{2}$, positive parity ($n S_{1/2}$) hydrogen.
The normalized wave functions read~\cite{landau:1982}:
\begin{align}
	\Psi_{n{1\over 2} m}(\vb{r}) & = \pmqty{ F_{n}(r) \sqrt{\frac{1}{4\pi}}\,\xi_m
\\
	G_{n}(r)\sqrt{\frac{3}{4\pi}} \bm{\sigma} \cdot \hat{\bf r} \,\xi_m},
\end{align}
where the Pauli spinors $\xi_{1\over 2}=\pmqty{1\\0}$ and $\xi_{-{1\over 2}}=\pmqty{0\\1}$.

Near the origin, $F_{n}(r)$ roughly scales as~\cite{Bethe1977}
\begin{align}
F_{n}(r)  \approx
R^{\rm Sch}_{n}(0) \left({2 r\over n a_0}\right)^{-\frac{Z^2\alpha^2}{2}},
\label{Fn:asym:small:r}
\end{align}
where $a_0= \hbar/(m c Z\alpha)$ is the Bohr radius, and
$R^{\rm Sch}_{n}(0)$ represents the radial Schr\"{o}dinger wave function at the origin for
the $nS$ hydrogen state.  We have also taken the nonrelativistic approximation
$\sqrt{1-Z^2\alpha^2}\approx 1-Z^2\alpha^2/2$ in the exponent.
The singularity has noticeable effect only when $r\lessapprox {n a_0\over 2} \exp(-2/Z^2\alpha^2)\sim
{n a_0\over 2} 10^{-16300/Z^2}$~\cite{Itzykson:1980rh},
which is even many orders shorter than the length scale related to the QED Landau pole!

A more careful expansion of the Dirac radial wave function near the origin
in first few powers of $Z\alpha$ reveals that
\begin{align}
& \lim_{r\to 0} F_{n}(r)= R^{\rm Sch}_{n}(0)\left(1-r/a_0\right)
\left(1-\frac{Z^2\alpha^2}{2} \ln r+\cdots\right).
\nn \\
\label{Fn:asym:small:r:fixed:order}
\end{align}
The wave function is manifestly logarithmically divergent
at the origin.

\paragraph{\color{blue} First-order quantum-mechanical perturbation theory.}
Let us first try to build some insight via a useful exercise
in the nonrelativistic quantum mechanics.
The electron in the hydrogen-like atom is undoubtedly nonrelativistic,
with $v/c\sim Z\alpha\sim\ll 1$). As can be found in many texts in
quantum mechanics~\cite{landau:1982,Holstein2014},
it is convenient to conduct a nonrelativistic reduction ($v/c$ expansion) to Dirac equation,
and end up with an effective Schr\"dinger equation:
\begin{align}
H_{\rm eff} \psi({\bf r})= E\psi({\bf r}),
\end{align}
where $\psi$ is a Pauli spinor.
The effective Hamiltonian is made of
\begin{align}
\label{H0:Coulomb}
 H_{\rm eff} = H_0 + \Delta H= H_0+ H_{\rm kin}+H_{\rm Dar}+H_{\rm s.o.}.
\end{align}
Here $H_0=-{\nabla^2\over 2m}- {e^2\over 4\pi r}$ is the nonrelativistic
Coulomb Hamiltonian (for brevity, we adopt the $\hbar=1$ unit from now on).
The kinetic ${\bf p}^4$ correction, Darwin term (responsible for the zitterbewegung motion) and
the spin-orbit terms comprise the ${\cal O}(1/c^2)$ relativistic corrections to $H_0$:
\begin{align}
& H_{\rm kin}= -{\bm{\nabla}^4 \over 8 m^3 c^2},\quad H_{\rm Dar}= -\frac{1}{8m^2 c^2} \nabla^2 \frac{Z\alpha}{r}.
\nn\\
& H_{\rm s.o.}= {Z\alpha\over 4m^2 c^2 r^3} \bm{\sigma}\cdot{\bf L}.
\label{DeltaH:perturb}
\end{align}

The fine structure of hydrogen level in \eqref{energy:level:hydrogen}
is readily recovered by incorporating \eqref{DeltaH:perturb} as small perturbations.
Here we explore the first-order perturbative correction to the Schr\"{o}dinger wave function
at the origin for the $1S$ state, again treating \eqref{DeltaH:perturb} as the perturbation.
We wonder whether the correction is finite or not.
Following the standard first-order perturbation theory in quantum mechanics, we have
\begin{align}
\Delta R^{(1)}_{10}(0)\sim
\int_0^{\infty}\! {dk\over 2\pi}\, R^{\rm Sch}_{k0}(0) {\langle k0\vert \Delta H\vert 10\rangle \over E_{10}-E_{k0}}.
\label{QM:pert:corr:wf:at:origin}
\end{align}
To our purpose, we need only consider the contribution from the continuum states
when summing over the intermediate states. Since $\Delta H$ is rotation scalar,
those surviving intermediates states must have $l=0$.

It turns out that the integration in \eqref{QM:pert:corr:wf:at:origin} are
dominated by the the high-frequency intermediate states.
To regularize the occurring UV divergence, we impose a hard momentum cutoff
$mZ\alpha \ll \Lambda \ll m$ in \eqref{QM:pert:corr:wf:at:origin}.
Accordingly, each piece of $\Delta H$ renders the following divergent
correction to the $1S$ wave function at the origin:
\bseq
\begin{align}
& \Delta R^{(1)}_{10}(0)\Big|_{\rm kin} = R^{\rm Sch}_{10}(0)\left( {Z\alpha\Lambda\over \pi m}+
Z^2\alpha^2 \ln \Lambda\right),
\label{Kin:corr:wf}
\\
& \Delta R^{(1)}_{10}(0)\Big|_{\rm Dar} =  R^{\rm Sch}_{10}(0)\left( -{Z\alpha\Lambda\over 2 \pi m}-
{Z^2\alpha^2\over 2} \ln \Lambda\right),
\label{Darwin:corr:wf}
\end{align}
\label{UV:Div:Corrs:to:WF:DeltaH}
\eseq
and $H_{\rm s.o.}$ yields a null correction for $S$-wave hydrogen.
Assuming $\Lambda \sim mZ\alpha$, one then verifies that each of $\Delta H$ does
generate a correction of order $v^2 \sim Z^2\alpha^2$, compatible with the $1/c^2$ scaling
in \eqref{DeltaH:perturb}.
Summing each entry in \eqref{UV:Div:Corrs:to:WF:DeltaH}, we find
\begin{align}
\Delta R^{(1)}_{10}(0)\Big|_{\rm \Delta H} = R^{\rm Sch}_{10}(0)\left({Z\alpha\Lambda\over 2 \pi m}+
{Z^2\alpha^2\over 2} \ln \Lambda\right).
\label{Total:corr:wf:at:the:origin}
\end{align}
Intriguingly, the ${Z^2\alpha^2\over 2} \ln \Lambda$ term coincides with what appears
in the expanded Dirac wave function \eqref{Fn:asym:small:r:fixed:order}, provided
$\Lambda$ is replaced with $1/r$. This might be a strong indicator
that the kinetic and Darwin terms are responsible for generating the mild logarithmic divergence
of the Dirac wave function near the origin.
The linear power divergence in \eqref{Total:corr:wf:at:the:origin} is clearly absent in the original
Dirac wave function \eqref{Fn:asym:small:r:fixed:order}. In our opinion,
these UV power divergences arise from our specific use of the
sharp momentum cutoff as the UV regulator, which may be viewed as an artifact.

\paragraph{\color{blue} OPE in relativistic QED fails.}
To reformulate the relativistic Dirac equation in an external Coulomb potential
in a field-theoretical fashion, one may naturally attempt to couple relativistic
QED with a external electrostatic field.
Analogous to our previous work~\cite{Huang:2018yyf,Huang:2018ils},
rather than introduce the classical electromagnetic field,
we introduce the heavy nucleus effective theory (HNET) (very similar to the famous
heavy quark effective theory~\cite{Eichten:1989zv,Georgi:1990um}),
and assign a HNET field $N$ to take into account the quantum fluctuation
around a positively-charged and static nucleus.
The relativistic UV theory tailored to describe atoms is then obtained
by combining the QED and HNET:
\begin{align}
{\mathcal L}_{\rm UV} = \bar{\Psi}(i\slashed{D}-m)\Psi +N^\dagger iD_0 N-\frac{1}{4}F_{\mu\nu}F^{\mu\nu},
\label{UV:atom:lagrangian}
\end{align}
where the covariant derivatives acting on electron and nucleus are
$D_{\mu}=\partial_{\mu}+ie A_{\mu}$ and $D_{\mu}=\partial_{\mu}-iZ eA_{\mu}$,
respectively. \eqref{Dirac:equation:Coulomb:potential} only concerns the $M_N\to \infty$ limit,
which is insensitive to the nucleus's detailed properties such as magnetic moment and charge radius.
We thereby retain only the leading HNET lagrangian in \eqref{UV:atom:lagrangian},
and treat $N$ as a bosonic field because the spin degree of freedom decouples.

Following \cite{Huang:2018yyf,Huang:2018ils}, we tentatively express the Dirac
wave function of a hydrogen atom in term of the following vacuum-to-atom matrix element:
\begin{align}
\Psi_{njm\Pi}({\bf r}) \equiv \langle 0 \vert  \Psi({\bf r}) N(0) \vert n j m;\Pi \rangle,
\label{Dirac:wf:nonlocal:matrix:element}
\end{align}
where $\Pi=\pm 1$ signifies the parity.
Note both field operators are defined in an equal time $t=0$.
This spatially-nonlocal matrix element clearly is not gauge invariant.

The universal near-origin behavior of the Dirac wave function may be
transparently understood from \eqref{Dirac:wf:nonlocal:matrix:element},
once the Wilson expansion of the operator product~\cite{Wilson:1969zs}
for $\phi({\bf x}) N(0)$ is conducted.
In the ${\bf r}\to 0$ limit, the anticipated OPE relation reads
\begin{align}
\Psi_R({\bf r}) N_R(0) =  \left(1+c(r)\right) [\Psi N]_R(0)+\cdots,
\label{OPE:QED:coordinate:space}
\end{align}
where $r\equiv |{\bf r}|$, the notion ``$[\ldots]_R$'' implies the renormalized operator,
and the ellipsis indicates the higher-dimensional operators carrying more derivatives.
Were the Wilson coefficient $c(r)$ exhibit the desired $Z^2\alpha^2 \ln r$ structure
as in \eqref{Fn:asym:small:r:fixed:order}, we would successfully accomplish our goal.

\newcommand{\QEDwidth}{1.cm}
\newcommand{\QEDheight}{1.3cm}
\newcommand{\QEDLOCALHeight}{0.4cm}
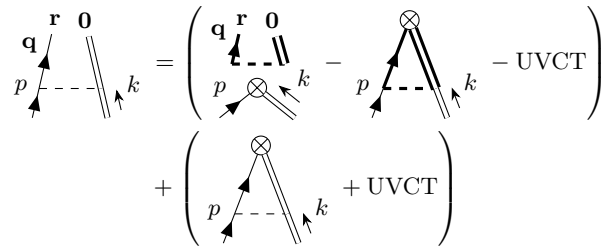
\begin{figure}[!hbtp]
	\centering
	\begin{displaymath}\begin{split}
		\begin{tikzpicture}[baseline=($(p1)!0.5!(x)$)]
			\begin{feynman}
				\vertex (p1);
				\vertex[right=\QEDwidth of p1] (p2);
				\vertex at ($(p1)!0.3!(p2)+(0,\QEDheight)$) (x) {\(\vb{r}\)};
				\vertex at ($(p1)!0.7!(p2)+(0,\QEDheight)$) (0) {\(\vb{0}\)};
				\vertex at ($(p1)!0.3!(x)$) (y1);
				\vertex at ($(p2)!0.3!(0)$) (z1);
				\diagram* {
				(y1) -- [scalar] (z1);
				(p1) -- [qed,edge label=\(p\)] (y1);
				(p2) -- [HQET,momentum'=\(k\)] (z1);
				(y1) -- [qed,edge label=\(\vb{q}\)] (x);
				(z1) -- [HQET] (0);
				};
			\end{feynman}
		\end{tikzpicture}&=
		\pqty{\begin{tikzpicture}[baseline=($(p1)!0.5!(x2)$)]
			\begin{feynman}
				\vertex                  (p1) ;
				\vertex[right=\QEDwidth of p1] (p2) ;
				\vertex at ($(p1)!0.3!(p2)+(0,\QEDheight)$) (x2) {\(\vb{r}\)};
				\vertex at ($(p1)!0.7!(p2)+(0,\QEDheight)$) (02) {\(\vb{0}\)};
				\node[crossed dot,minimum size=\dotsize] at ($(p1)!0.5!(p2)+(0,\QEDLOCALHeight)$) (x) ;
				\vertex at ($(p1)!0.1!(x2)$) (p12);
				\vertex[right=\QEDwidth of p12] (p22);
				\vertex at ($(p12)!0.5!(x2)$) (y12);
				\vertex at ($(p22)!0.5!(02)$) (z12);
				\diagram* {
					(p1) -- [qed,edge label=\(p\)] (x);
					(p2) -- [HQET,momentum'=\(k\)] (x);
				};
				\diagram* {
				(y12) -- [very thick, scalar] (z12);
				(y12) -- [very thick,qed,edge label=\(\vb{q}\)] (x2);
				(z12) -- [very thick, HQET] (02);
				};
			\end{feynman}
		\end{tikzpicture}-\begin{tikzpicture}[baseline=($(p1)!0.5!(x)$)]
			\begin{feynman}
				\vertex (p1);
				\vertex[right=\QEDwidth of p1] (p2);
				\node[crossed dot,minimum size=\dotsize] at ($(p1)!0.5!(p2)+(0,\QEDheight)$) (x) ;
				\vertex at ($(p1)!0.3!(x)$) (y1);
				\vertex at ($(p2)!0.3!(x)$) (z1);
				\diagram* {
				(y1) -- [very thick,scalar] (z1);
				(p1) -- [qed,edge label=\(p\)] (y1);
				(p2) -- [HQET,momentum'=\(k\)] (z1);
				(y1) -- [very thick,qed] (x);
				(z1) -- [very thick,HQET] (x);
				};
			\end{feynman}
		\end{tikzpicture}-\text{UVCT}}\\&+\pqty{
		\begin{tikzpicture}[baseline=($(p1)!0.5!(x)$)]
			\begin{feynman}
				\vertex (p1);
				\vertex[right=\QEDwidth of p1] (p2);
				\node[crossed dot,minimum size=\dotsize] at ($(p1)!0.5!(p2)+(0,\QEDheight)$) (x) ;
				\vertex at ($(p1)!0.3!(x)$) (y1);
				\vertex at ($(p2)!0.3!(x)$) (z1);
				\diagram* {
				(y1) -- [scalar] (z1);
				(p1) -- [qed,edge label=\(p\)] (y1);
				(p2) -- [HQET,momentum'=\(k\)] (z1);
				(y1) -- [qed] (x);
				(z1) -- [HQET] (x);
				};
			\end{feynman}
		\end{tikzpicture}+\text{UVCT}}
	\end{split}\end{displaymath}
\caption{\label{Fig:NLO:OPE:coordinate:space} The next-to-leading Wilson expansion for the product of
the $\Psi$ and $N$ fields. The original diagram is reexpressed as the sum of two terms,
by subtracting and adding the renormalized Green function inserted with a local $\Psi N$ operator.}
\end{figure}

The renormalization of $\Psi N$ resembles very much with that of the
heavy-light quark current $\bar{q}\Gamma h_v$ in HQET~\cite{Manohar:2000dt},
where the logarithmic UV divergence is first encountered at order $Z\alpha$.
This immediately indicates the Wilson coefficient should exhibit $Z\alpha\ln r$ structure.
Following the instruction expounded in \cite{Huang:2018yyf},
by analyzing the asymptotic behavior of the four-point Green function in the $r\to 0$ limit
as guided by Fig.~\ref{Fig:NLO:OPE:coordinate:space}, one finds that $c(r)$ at order-$Z\alpha$ reads
\begin{align}
	c(r)=\begin{dcases}
		{Z \alpha  \over 2 \pi} \ln r+\cdots, \quad\text{Feynman}
\\
		{Z \alpha  \over \pi} \ln r+\cdots,   \quad \text{Coulomb}
	\end{dcases}
\label{OPE:QED:coordinate:space}
\end{align}
in Feynman and Coulomb gauge, respectively.

Notwithstanding its gauge dependence, the Wilson coefficient $c(r)$ in
\eqref{OPE:QED:coordinate:space} does contain the $\ln r$ term, which unfortunately arises prematurely at order $Z\alpha$,
rather than the desired $Z^2\alpha^2\ln r$ in \eqref{Fn:asym:small:r:fixed:order}.

Therefore, we have to admit that the Wilson expansion formulated
in our UV theory for atoms, \eqref{UV:atom:lagrangian},
fails to account for the near-the-origin behavior of the hydrogen Dirac wave function.
This failure may cast some doubt on the legitimacy of interpreting the matrix element
\eqref{Dirac:wf:nonlocal:matrix:element} as the Dirac wave function.
The very short distance ($r\ll 1/m$) behavior of the Dirac wave function
cannot be correctly accounted by the OPE based on \eqref{UV:atom:lagrangian},
which may hint that one should not associate physical significance
with the Dirac wave function literally as $r\to 0$.

\paragraph{\color{blue} NREFT lagrangian and the desired OPE.}
Inspired by the success of quantum mechanical perturbation theory
encoded in \eqref{Total:corr:wf:at:the:origin}, one might like to reexamine the OPE
formulated in the nonrelativstic EFT rather than the UV theory \eqref{UV:atom:lagrangian}, but
including a certain class of relativistic corrections.

A natural option is to replace the QED lagrangian in \eqref{UV:atom:lagrangian} by the
nonrelativistic QED (NRQED)~\cite{Caswell:1985ui}.
To mimic the dynamics underpinning the Coulomb solution of the
Dirac equation, we deliberately drop all the occurrences of the $\bf A$ field.
In the lowest order of $1/M_N$ and through the order of $v^2/c^2$,
the nonrelativistic EFT for atoms~\cite{Pineda:1997ie} then reduces to
\begin{align}
& {\mathcal L}_{\rm NREFT}=\psi^\dagger \Bigg\{iD_0+\frac{\bm{\nabla}^2}{2m}+\frac{\bm{\nabla}^4}{8m^3}                                                      
\nn\\
& -c_D e\frac{\bm{\nabla}^2A^0}{8m^2}+ic_Se\frac{ \bm{\sigma}\cdot\pqty{\bm{\nabla}A^0 \cross \bm{\nabla}}}{4m^2}+\cdots\Bigg\}\psi
\label{Lag:NREFT:without:vec:A}\\
& +N^\dagger i D^0 N+\frac{c_4}{m^2}\psi^\dagger\psi N^\dagger N+
{1\over 2} \left(\nabla A^0\right)^2+\cdots.\nn
\end{align}
The positron sector in NRQED has been neglected.
At tree level, the short-distance coefficients associated with the Darwin
and spin-orbit terms are simply $c_D=c_S=1$.
Note that an intrinsic UV cutoff $\Lambda \leq m$ must be imposed
to make \eqref{Lag:NREFT:without:vec:A} meaningful.

Eq.~\eqref{Lag:NREFT:without:vec:A} is no longer gauge invariant.
However, this does not necessarily imply a nuisance. In fact,
Coulomb gauge $\nabla \cdot {\bf A}=0$ is a convenient choice for tackling
nonrelativistic bound state problem,
where the effects of $\bf A$ and $A^0$ can be cleanly separated.
Unlike the physics underpinning the Lamb shift~\cite{Pineda:1997ie},
Dirac equation \eqref{Dirac:equation:Coulomb:potential} {\it per se}
does not involve the effect of the physical
photon excited by the $\bf A$ field.
The $A^0$ field is the very agent to mediate the instantaneous Coulomb interaction,
which provides the crucial binding mechanism all atoms.
We will stay exclusively with the Coulomb gauge.

One can descend from \eqref{UV:atom:lagrangian} to \eqref{Lag:NREFT:without:vec:A}
via the perturbative matching procedure.
There may potentially arise a contact interaction in NREFT due to integration out of the shortly-lived
$e^+e^-$ intermediate states. Nevertheless, through an explicit one-loop matching calculation,
we find no room for the contact interaction to contribute. Thus we conclude $c_4=\mathcal{O}(Z^3\alpha^3)$.

Due to the single-pole structure of the nonrelativistic propagators
and the instantaneous Coulomb interaction, self-energy diagrams for the $\varphi$ and $N$ fields, as well
as crossed ladder diagrams simply vanish in arbitrarily high order.
As a gratifying simplification, we only need consider the uncrossed Coulomb ladder diagrams in
NREFT sector.

Analogous to \eqref{Dirac:wf:nonlocal:matrix:element}, we tentatively define the large component of the
Dirac wave function for the $nS_{1/2}$ as the following nonlocal matrix element:
\begin{align}
F_{n}(r) {1\over \sqrt{4\pi}}\,\xi_m
= \langle 0 \vert \psi({\bf r}) N(0) \vert n S_{1/2},m \rangle,
\label{Dirac:wf:nonlocal:matrix:element:NREFT:field}
\end{align}
with the Dirac field $\Psi$ in \eqref{Dirac:wf:nonlocal:matrix:element}
replaced by the Pauli spinor field $\psi$.

In \cite{Huang:2018ils}, we have successfully accounted for the
near-the-origin behavior of the Klein-Gordon hydrogen wave function
with the aid of OPE formulated in the NREFT.
Hearteningly, the universal near-the-origin behavior of the Dirac wave function can
also be understood by applying the OPE to
\eqref{Dirac:wf:nonlocal:matrix:element:NREFT:field}.
In the small $r$ limit, the OPE relation turns out to read
\begin{align}
& \lim_{r \to {1\over m}} \psi({\bf r}) N(0) =  {\mathcal C}(r) [\psi N]_R(0)+\cdots,
\label{OPE:NREFT:rel:corr}
\\
& {\mathcal C}(r) = 1- mZ\alpha r -{Z^2\alpha^2\over 2} \left(\ln \mu r+{\rm const} \right)
+{\cal O}(Z^3\alpha^3).
\nn
\end{align}
This formula constitutes the key result of this work.
The ${\cal O}(Z\alpha)$ Wilson coefficient was first derived in
the nonrelativistic Schr\"{o}dinger EFT~\cite{Huang:2018yyf},
which is invoked to explain the universal electron-nucleus
coalescence behavior of the atomic Schrodinger wave function~\cite{Loewdin1954,Kato1957}.

The ${\cal O}(Z^2\alpha^2)$ coefficient in \eqref{OPE:NREFT:rel:corr}
is highly desirable, which reproduces the logarithmic scaling of the
Dirac wave function at small $r$ in \eqref{Fn:asym:small:r:fixed:order}.
Moreover, it also depends on an artificial scale $mZ\alpha < \mu \le m$,
which is the renormalization scale associated with the local operator $[\psi N]_R$.

In \cite{Huang:2018ils}, we have shown that the kinetic ${\bf p}^4$ term plays the vital role
in generating the $Z^2\alpha^2\ln r$ coefficient for spin-0 electron.
For spin-${1\over 2}$ electron, the Darwin term is as important as the kinetic term, both of which
represent the ${\cal O}(1/c^2)$ relativistic corrections in \eqref{Lag:NREFT:without:vec:A} (Spin-orbit
term yields a vanishing contribution).
The extra contribution from Darwin term in the Dirac case
explains why the $Z^2\alpha^2\ln r$ coefficient in \eqref{OPE:NREFT:rel:corr} differs by the factor of two
from its counterpart in the Klein-Gordon case.

The OPE relation in the momentum space can be obtained by
Fourier transforming \eqref{OPE:NREFT:rel:corr}.
In the ${\bf q}\to m$ limit, one has
\begin{align}
& \widetilde{\psi}({\bf q}) N(0) \equiv
\int\!\! d^3{\bf{r}}\, e^{-i{\bf{q}}\cdot {\bf{r}}}\varphi({\bf r}) N(0)
\nn\\
& \qquad \qquad \; \to  \widetilde{\cal C}(q)[\psi N]_R(0)+\cdots,
\label{OPE:NREFT:momentum:space}\\
& \widetilde{\mathcal C}(q) = {8\pi mZ\alpha+{\cal O}(Z^3\alpha^3) \over {\bf q}^4}
 - {\pi^2 Z^2\alpha^2+{\cal O}(Z^4\alpha^4) \over |{\bf q}|^3}, \nn
\end{align}
where we have suppressed the free-theory
Wilson coefficient $(2\pi)^3\delta^{(3)}({\bf q})$.
In contrast to the renormalizable field theory, here the coexistence of
both ${\bf q}^{-4}$ and $|{\bf q}|^{-3}$ scalings in
the Wilson coefficient is peculiar in NREFT, which is dubbed
the {\it double-layer form of OPE}~\cite{Huang:2018ils}.

\paragraph{\color{blue} Renormalization of local Operator in NREFT.}
To better understand why the $Z^2\alpha^2\ln r$ arises in the Wilson coefficient,
let us first examine the renormalized composite operator $\psi N$, which
appears in the right hand side of OPE in \eqref{OPE:NREFT:rel:corr}.

It is known that the logarithmic UV singularity arises first at two-loop order
in NRQED (NRQCD) vertex diagrams, first discovered by matching the vector current
from QCD onto NRQCD~\cite{Czarnecki:1997vz,Beneke:1997jm}.
Later an explicit two-loop calculation is also conducted in the nonrelativistic EFT of QCD~\cite{Luke:1999kz}.
It is a certain class of relativistic effects that play a vital role in generating logarithmic UV divergence.
This knowledge hints that the local operator $\psi N$ may well be subject to a
multiplicative renormalization at order $Z^2\alpha^2$.

\begin{figure}[!hbtp]
	\centering
	\begin{tabular}{ccccc}
		  & \null\hfill
		\subfloat[]{
			$$\begin{tikzpicture}[baseline=($(p1)!0.5!(x)$)]
					\begin{feynman}
						\vertex (p1);
						\vertex[right=1.7cm of p1] (p2);
						\node[crossed dot] at ($(p1)!0.5!(p2)+(0,2cm)$) (x) ;
						\vertex at ($(p1)!0.2!(x)$) (y1);
						\vertex at ($(p2)!0.2!(x)$) (z1);
						\vertex at ($(p1)!0.6!(x)$) (y2);
						\vertex at ($(p2)!0.6!(x)$) (z2);
						\vertex at ($(y1)!0.5!(z1)$) (t);
						\diagram* {
						(y1) -- [scalar] (z1);
						(y2) -- [scalar] (z2);
						(p1) -- [fermion] (y1);
						(p2) -- [HQET] (z1);
						(y1) -- [insertion={[style=thick,size=3pt]0.7},with arrow=0.4] (y2);
						(z1) -- [HQET] (z2);
						(y2) -- [fermion] (x);
						(z2) -- [HQET] (x);
						};
					\end{feynman}
				\end{tikzpicture}$$
		}\hfill
		  & \subfloat[]{
			$$\begin{tikzpicture}[baseline=($(p1)!0.5!(x)$)]
					\begin{feynman}
						\vertex (p1);
						\vertex[right=1.7cm of p1] (p2);
						\node[crossed dot] at ($(p1)!0.5!(p2)+(0,2cm)$) (x) ;
						\vertex at ($(p1)!0.2!(x)$) (y1);
						\vertex at ($(p2)!0.2!(x)$) (z1);
						\vertex at ($(p1)!0.6!(x)$) (y2);
						\vertex at ($(p2)!0.6!(x)$) (z2);
						\vertex at ($(y1)!0.5!(z1)$) (t);
						\diagram* {
						(y1) -- [scalar] (z1);
						(y2) -- [scalar] (z2);
						(p1) -- [fermion] (y1);
						(p2) -- [HQET] (z1);
						(y1) -- [fermion] (y2);
						(z1) -- [HQET] (z2);
						(y2) -- [insertion={[style=thick,size=3pt]0.7},with arrow=0.4] (x);
						(z2) -- [HQET] (x);
						};
					\end{feynman}
				\end{tikzpicture}$$
		}
		\hfill
		  & \subfloat[]{
			$$\begin{tikzpicture}[baseline=($(p1)!0.5!(x)$)]
					\begin{feynman}
						\vertex (p1);
						\vertex[right=1.7cm of p1] (p2);
						\node[crossed dot] at ($(p1)!0.5!(p2)+(0,2cm)$) (x) ;
						\vertex at ($(p1)!0.2!(x)$) (y1);
						\node [square dot,minimum size=1.3mm] at ($(y1)$) (tmp);
						\vertex at ($(p2)!0.2!(x)$) (z1);
						\vertex at ($(p1)!0.6!(x)$) (y2);
						\vertex at ($(p2)!0.6!(x)$) (z2);
						\diagram* {
						(y1) -- [scalar] (z1);
						(y2) -- [scalar] (z2);
						(p1) -- [fermion] (y1);
						(p2) -- [HQET] (z1);
						(y1) -- [fermion] (y2);
						(z1) -- [HQET] (z2);
						(y2) -- [fermion] (x);
						(z2) -- [HQET] (x);
						};
					\end{feynman}
				\end{tikzpicture}$$
		}
		\hfill
		  & \subfloat[]{
			$$\begin{tikzpicture}[baseline=($(p1)!0.5!(x)$)]
					\begin{feynman}
						\vertex (p1);
						\vertex[right=1.7cm of p1] (p2);
						\node[crossed dot] at ($(p1)!0.5!(p2)+(0,2cm)$) (x) ;
						\vertex at ($(p1)!0.2!(x)$) (y1);
						\vertex at ($(p2)!0.2!(x)$) (z1);
						\vertex at ($(p1)!0.6!(x)$) (y2);
						\node [square dot,minimum size=1.3mm] at ($(y2)$) (tmp);
						\vertex at ($(p2)!0.6!(x)$) (z2);
						\diagram* {
						(y1) -- [scalar] (z1);
						(y2) -- [scalar] (z2);
						(p1) -- [fermion] (y1);
						(p2) -- [HQET] (z1);
						(y1) -- [fermion] (y2);
						(z1) -- [HQET] (z2);
						(y2) -- [fermion] (x);
						(z2) -- [HQET] (x);
						};
					\end{feynman}
				\end{tikzpicture}$$
		}\\
		  & \null\hfill
		\subfloat[]{
			$$\begin{tikzpicture}[baseline=($(p1)!0.5!(x)$),remember picture]
					\begin{feynman}
						\vertex (p1);
						\vertex[right=1.7cm of p1] (p2);
						\node[crossed dot] at ($(p1)!0.5!(p2)+(0,2cm)$) (x) ;
						\vertex at ($(p1)!0.2!(x)$) (y1);
						\vertex at ($(p2)!0.2!(x)$) (z1);
						\vertex at ($(p1)!0.6!(x)$) (y2);
						\vertex at ($(p2)!0.6!(x)$) (z2);
						\diagram* {
						(y1) -- [scalar] (z1);
						(y2) -- [scalar] (z2);
						(p1) -- [fermion] (y1);
						(p2) -- [HQET] (z1);
						(y1) -- [fermion] (y2);
						(z1) -- [HQET] (z2);
						(y2) -- [fermion] (x);
						(z2) -- [HQET] (x);
						};
						\node [square dot] at ($(y1)$) (tmp);
						\node [empty square dot] at ($(y1)$) (tmp1);
					\end{feynman}
				\end{tikzpicture}$$
			\label{subfig:SO1}
		}\hfill
		  & \subfloat[]{
			$$\begin{tikzpicture}[baseline=($(p1)!0.5!(x)$),remember picture]
					\begin{feynman}
						\vertex (p1);
						\vertex[right=1.7cm of p1] (p2);
						\node[crossed dot] at ($(p1)!0.5!(p2)+(0,2cm)$) (x) ;
						\vertex at ($(p1)!0.2!(x)$) (y1);
						\vertex at ($(p2)!0.2!(x)$) (z1);
						\vertex at ($(p1)!0.6!(x)$) (y2);
						\vertex at ($(p2)!0.6!(x)$) (z2);
						\diagram* {
						(y1) -- [scalar] (z1);
						(y2) -- [scalar] (z2);
						(p1) -- [fermion] (y1);
						(p2) -- [HQET] (z1);
						(y1) -- [fermion] (y2);
						(z1) -- [HQET] (z2);
						(y2) -- [fermion] (x);
						(z2) -- [HQET] (x);
						};
						\node [square dot] at ($(y2)$) (tmp);
						\node [empty square dot] at ($(y2)$) (tmp1);
					\end{feynman}
				\end{tikzpicture}$$
			\label{subfig:SO2}
		}\hfill
		& \subfloat[]{
			\begin{tikzpicture}[baseline=($(p1)!0.5!(x)$)]
				\begin{feynman}
					\vertex (p1);
					\vertex[right=1.7cm of p1] (p2);
					\node[crossed dot] at ($(p1)!0.5!(p2)+(0,2cm)$) (x) ;
					\vertex at ($(p1)!0.5!(p2)$) (p0) ;
					\node[empty dot] at ($(p0)!0.4!(x)$) (i);
					\vertex at ($(p0)!0.66!(x)+(0.85cm,0)$) (y1);
					\vertex[right=2.2cm of y1] (z1);
					%
					\diagram* {
					(p1) -- [fermion] (i);
					(p2) -- [HQET] (i);
					(i) -- [fermion,half left] (x);
					(i) -- [HQET, half right] (x);
					};
				\end{feynman}
			\end{tikzpicture}
		}\hfill
		& \subfloat[]{
			\begin{tikzpicture}[baseline=($(p1)!0.5!(x)$)]
				\begin{feynman}
					\vertex (p1);
					\vertex[right=1.7cm of p1] (p2);
					\node[crossed dot] at ($(p1)!0.5!(p2)+(0,2cm)$) (x) ;
					\vertex at ($(p1)!0.5!(p2)$) (p0) ;
					\node[empty dot] at ($(p0)!0.4!(x)$) (i);
					\vertex at ($(p0)!0.66!(x)+(-0.65cm,0)$) (y1);
					\vertex at ($(p0)!0.66!(x)+(0.65cm,0)$) (z1);
					%
					\diagram* {
					(p1) -- [fermion] (i);
					(p2) -- [double distance=1pt] (i);
					(i) -- [fermion,quarter left] (y1);
					(y1) -- [fermion, quarter left] (x);
					(i) -- [HQET, quarter right] (z1);
					(z1) -- [HQET,quarter right] (x);
					(y1) -- [scalar] (z1);
					};
				\end{feynman}
			\end{tikzpicture}
		}
\end{tabular}
\caption{\label{Fig:vertex:corr:all} Representative diagrams for local operator renormalization. The cap represents the
insertion of the operator $\psi N$, cross stands for the ${\bf p}^4$ relativistic correction,
solid square for the Darwin vertex, while empty square for spin-orbital vertex, which making vanishing contribution
in this case. The empty circle represents the contact interaction.
The last two diagrams are beyond the prescribed accuracy of ${\cal O}(Z^2\alpha^2)$.}
\end{figure}
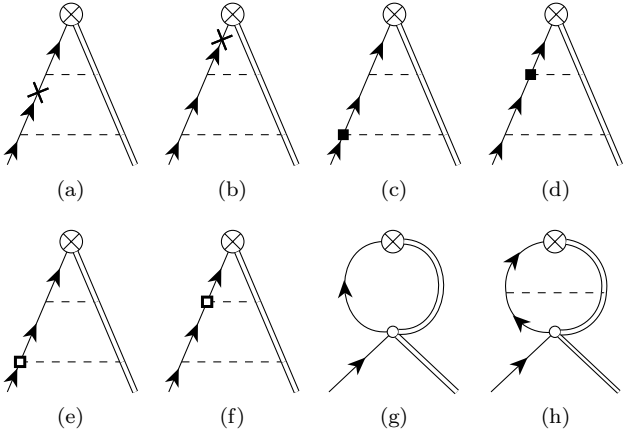

We use dimensional regularization (spacetime dimension $d=4-\epsilon$) to regularize UV divergence,
which respects the all the cherished symmetry. Power divergences are automatically absent in this scheme.
At order $Z^2\alpha^2$, only two diagrams in Fig.~\ref{Fig:vertex:corr:all} generate logarithmic UV divergence:
\bseq
\begin{align}
& {\cal G}_{\ref{Fig:vertex:corr:all}a)}= {Z^2\alpha^2\over 2} \left({1\over \epsilon}+\ln \mu^2 +{\rm finite}\right),
\\
& {\cal G}_{\ref{Fig:vertex:corr:all}c)} = -{Z^2\alpha^2\over 4} \left({1\over\epsilon}+\ln\mu^2+{\rm finite} \right),
\end{align}
\eseq
The first term comes from Fig.~\ref{Fig:vertex:corr:all}$a)$ which
involves a ${\bf p}^4$ insertion in the bottom loop~\cite{Huang:2018yyf}.
The second term is new, which arising from Fig.~\ref{Fig:vertex:corr:all}$a)$ containing
a Darwin vertex. The divergence of order-$v^0$ is specific to NRQED,
because the Coulomb enhancement factor $\propto (1/v)^2$ compensates the relativistic
correction $\propto v^2$.

The total UV divergence at two loop order is then is $Z^2\alpha ^2 {1\over 4 \epsilon}$.
Define operator renormalization constant via $\bqty{\psi N}_R=Z_\mathcal{S}{\psi N}$,
whose value in the MS scheme is then
\begin{align}
Z_{\mathcal{S}}=1-\frac{Z^2\alpha^2}{4\epsilon}+\cdots.
\label{compo-renorm}
\end{align}
The anomalous dimension of the operator $\psi N$ then reads
\begin{align}
\gamma_{\mathcal{S}}\equiv {d\ln Z_\mathcal{S} \over d\ln\mu}= {Z^2\alpha^2\over 2}.
\label{ano-dim}
\end{align}

	\onecolumngrid

	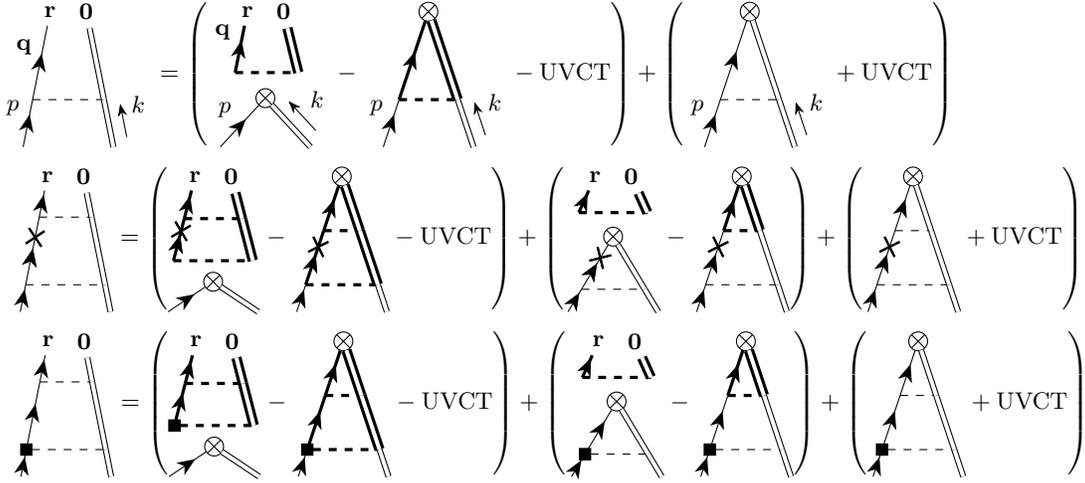
\begin{figure}[pbth]
		\centering
		\begin{displaymath}\begin{split}&\hspace{-2mm}
		\begin{tikzpicture}[baseline=($(p1)!0.5!(x)$)]
			\begin{feynman}
				\vertex (p1);
				\vertex[right=1.2cm of p1] (p2);
				\vertex at ($(p1)!0.3!(p2)+(0,1.8cm)$) (x) {\(\vb{r}\)};
				\vertex at ($(p1)!0.7!(p2)+(0,1.8cm)$) (0) {\(\vb{0}\)};
				\vertex at ($(p1)!0.35!(x)$) (y1);
				\vertex at ($(p2)!0.35!(0)$) (z1);
				\diagram* {
				(y1) -- [scalar] (z1);
				(p1) -- [fermion,edge label=\(p\)] (y1);
				(p2) -- [HQET,momentum'=\(k\)] (z1);
				(y1) -- [fermion,edge label=\(\vb{q}\)] (x);
				(z1) -- [HQET] (0);
				};
			\end{feynman}
		\end{tikzpicture}=
		\pqty{\begin{tikzpicture}[baseline=($(p1)!0.5!(x2)$)]
			\begin{feynman}
				\vertex                  (p1) ;
				\vertex[right=1.2cm of p1] (p2) ;
				\vertex at ($(p1)!0.3!(p2)+(0,1.8cm)$) (x2) {\(\vb{r}\)};
				\vertex at ($(p1)!0.7!(p2)+(0,1.8cm)$) (02) {\(\vb{0}\)};
				\node[crossed dot,minimum size=\dotsize] at ($(p1)!0.5!(p2)+(0,0.65cm)$) (x) ;
				\vertex at ($(p1)!0.1!(x2)$) (p12);
				\vertex[right=1.2cm of p12] (p22);
				\vertex at ($(p12)!0.5!(x2)$) (y12);
				\vertex at ($(p22)!0.5!(02)$) (z12);
				\diagram* {
					(p1) -- [fermion,edge label=\(p\)] (x);
					(p2) -- [HQET,momentum'=\(k\)] (x);
				};
				\diagram* {
				(y12) -- [very thick, scalar] (z12);
				(y12) -- [very thick,fermion,edge label=\(\vb{q}\)] (x2);
				(z12) -- [very thick, HQET] (02);
				};
			\end{feynman}
		\end{tikzpicture}-\begin{tikzpicture}[baseline=($(p1)!0.5!(x)$)]
			\begin{feynman}
				\vertex (p1);
				\vertex[right=1.2cm of p1] (p2);
				\node[crossed dot,minimum size=\dotsize] at ($(p1)!0.5!(p2)+(0,1.8cm)$) (x) ;
				\vertex at ($(p1)!0.35!(x)$) (y1);
				\vertex at ($(p2)!0.35!(x)$) (z1);
				\diagram* {
				(y1) -- [very thick,scalar] (z1);
				(p1) -- [fermion,edge label=\(p\)] (y1);
				(p2) -- [HQET,momentum'=\(k\)] (z1);
				(y1) -- [very thick,fermion] (x);
				(z1) -- [very thick,HQET] (x);
				};
			\end{feynman}
		\end{tikzpicture}-\text{UVCT}}
		+\pqty{
		\begin{tikzpicture}[baseline=($(p1)!0.5!(x)$)]
			\begin{feynman}
				\vertex (p1);
				\vertex[right=1.2cm of p1] (p2);
				\node[crossed dot,minimum size=\dotsize] at ($(p1)!0.5!(p2)+(0,1.8cm)$) (x) ;
				\vertex at ($(p1)!0.35!(x)$) (y1);
				\vertex at ($(p2)!0.35!(x)$) (z1);
				\diagram* {
				(y1) -- [scalar] (z1);
				(p1) -- [fermion,edge label=\(p\)] (y1);
				(p2) -- [HQET,momentum'=\(k\)] (z1);
				(y1) -- [fermion] (x);
				(z1) -- [HQET] (x);
				};
			\end{feynman}
		\end{tikzpicture}+\text{UVCT}}\\&
		\begin{tikzpicture}[baseline=($(p1)!0.5!(x)$)]
			\begin{feynman}
				\vertex (p1);
				\vertex[right=1.2cm of p1] (p2);
				\vertex at ($(p1)!0.3!(p2)+(0,1.8cm)$) (x) {\(\vb{r}\)};
				\vertex at ($(p1)!0.7!(p2)+(0,1.8cm)$) (0) {\(\vb{0}\)};
				\vertex at ($(p1)!0.2!(x)$) (y1);
				\vertex at ($(p2)!0.2!(0)$) (z1);
				\vertex at ($(p1)!0.7!(x)$) (y2);
				\vertex at ($(p2)!0.7!(0)$) (z2);
				\diagram* {
				(y1) -- [scalar] (z1);
				(y2) -- [scalar] (z2);
				(p1) -- [fermion] (y1);
				(p2) -- [HQET] (z1);
				(y1) -- [with arrow=0.4,insertion={[style=thick,size=3pt]0.7}] (y2);
				(z1) -- [HQET] (z2);
				(y2) -- [fermion] (x);
				(z2) -- [HQET] (0);
				};
			\end{feynman}
		\end{tikzpicture}=
		\pqty{\begin{tikzpicture}[baseline=($(p1)!0.5!(x2)$)]
			\begin{feynman}
				\vertex                  (p1) ;
				\vertex[right=1.2cm of p1] (p2) ;
				\node[crossed dot,minimum size=\dotsize] at ($(p1)!0.5!(p2)+(0,0.4cm)$) (x) ;
				\vertex[above=0.4cm of p1] (p12);
				\vertex[right=1.2cm of p12] (p22);
				\vertex at ($(p12)!0.3!(p22)+(0,1.4cm)$) (x2) {\(\vb{r}\)};
				\vertex at ($(p12)!0.7!(p22)+(0,1.4cm)$) (02) {\(\vb{0}\)};
				\vertex at ($(p12)!0.2!(x2)$) (y12);
				\vertex at ($(p22)!0.2!(02)$) (z12);
				\vertex at ($(p12)!0.6!(x2)$) (y22);
				\vertex at ($(p22)!0.6!(02)$) (z22);
				\diagram* {
				(p1) -- [fermion] (x);
				(p2) -- [HQET] (x);
				};
				\diagram* {
				(y12) -- [very thick, scalar] (z12);
				(y22) -- [very thick, scalar] (z22);
				(y12) -- [very thick,with arrow=0.4,insertion={[style=thick,size=3pt]0.7}] (y22);
				(z12) -- [very thick, HQET] (z22);
				(y22) -- [very thick,fermion] (x2);
				(z22) -- [very thick, HQET] (02);
				};
			\end{feynman}
		\end{tikzpicture}-\begin{tikzpicture}[baseline=($(p1)!0.5!(x)$)]
			\begin{feynman}
				\vertex (p1);
				\vertex[right=1.2cm of p1] (p2);
				\node[crossed dot,minimum size=\dotsize] at ($(p1)!0.5!(p2)+(0,1.8cm)$) (x) ;
				\vertex at ($(p1)!0.2!(x)$) (y1);
				\vertex at ($(p2)!0.2!(x)$) (z1);
				\vertex at ($(p1)!0.6!(x)$) (y2);
				\vertex at ($(p2)!0.6!(x)$) (z2);
				\vertex at ($(y1)!0.5!(z1)$) (t);
				\diagram* {
				(y1) -- [very thick,scalar] (z1);
				(y2) -- [very thick,scalar] (z2);
				(p1) -- [fermion] (y1);
				(p2) -- [HQET] (z1);
				(y1) -- [very thick,with arrow=0.4,insertion={[style=thick,size=3pt]0.7}] (y2);
				(z1) -- [very thick,HQET] (z2);
				(y2) -- [very thick,fermion] (x);
				(z2) -- [very thick,HQET] (x);
				};
			\end{feynman}
		\end{tikzpicture}-\text{UVCT}}
		+\pqty{\begin{tikzpicture}[baseline=($(p1)!0.5!(x2)$)]
			\begin{feynman}
				\vertex                  (p1) ;
				\vertex[right=1.2cm of p1] (p2) ;
				\node[crossed dot,minimum size=\dotsize] at ($(p1)!0.5!(p2)+(0,1cm)$) (x) ;
				\vertex at ($(p1)!0.3!(x)$) (y1);
				\vertex at ($(p2)!0.3!(x)$) (z1);
				\vertex[above=1cm of p1] (p12);
				\vertex[right=1.2cm of p12] (p22);
				\vertex at ($(p12)!0.3!(p22)+(0,0.8cm)$) (x2) {\(\vb{r}\)};
				\vertex at ($(p12)!0.7!(p22)+(0,0.8cm)$) (02) {\(\vb{0}\)};
				\vertex at ($(p12)!0.4!(x2)$) (y12);
				\vertex at ($(p22)!0.4!(02)$) (z12);
				\diagram* {
					(y1) -- [scalar] (z1);
					(p1) -- [fermion] (y1);
					(p2) -- [HQET] (z1);
					(y1) -- [with arrow=0.4,insertion={[style=thick,size=3pt]0.7}] (x);
					(z1) -- [HQET] (x);
				};
				\diagram* {
				(y12) -- [very thick, scalar] (z12);
				(y12) -- [very thick,fermion] (x2);
				(z12) -- [very thick, HQET] (02);
				};
			\end{feynman}
		\end{tikzpicture}-\begin{tikzpicture}[baseline=($(p1)!0.5!(x)$)]
			\begin{feynman}
				\vertex (p1);
				\vertex[right=1.2cm of p1] (p2);
				\node[crossed dot,minimum size=\dotsize] at ($(p1)!0.5!(p2)+(0,1.8cm)$) (x) ;
				\vertex at ($(p1)!0.2!(x)$) (y1);
				\vertex at ($(p2)!0.2!(x)$) (z1);
				\vertex at ($(p1)!0.6!(x)$) (y2);
				\vertex at ($(p2)!0.6!(x)$) (z2);
				\vertex at ($(y1)!0.5!(z1)$) (t);
				\diagram* {
				(y1) -- [scalar] (z1);
				(y2) -- [very thick,scalar] (z2);
				(p1) -- [fermion] (y1);
				(p2) -- [HQET] (z1);
				(y1) -- [with arrow=0.4,insertion={[style=thick,size=3pt]0.7}] (y2);
				(z1) -- [HQET] (z2);
				(y2) -- [very thick,fermion] (x);
				(z2) -- [very thick,HQET] (x);
				};
			\end{feynman}
		\end{tikzpicture}}+
		\pqty{\begin{tikzpicture}[baseline=($(p1)!0.5!(x)$)]
			\begin{feynman}
				\vertex (p1);
				\vertex[right=1.2cm of p1] (p2);
				\node[crossed dot,minimum size=\dotsize] at ($(p1)!0.5!(p2)+(0,1.8cm)$) (x) ;
				\vertex at ($(p1)!0.2!(x)$) (y1);
				\vertex at ($(p2)!0.2!(x)$) (z1);
				\vertex at ($(p1)!0.6!(x)$) (y2);
				\vertex at ($(p2)!0.6!(x)$) (z2);
				\vertex at ($(y1)!0.5!(z1)$) (t);
				\diagram* {
				(y1) -- [scalar] (z1);
				(y2) -- [scalar] (z2);
				(p1) -- [fermion] (y1);
				(p2) -- [HQET] (z1);
				(y1) -- [with arrow=0.4,insertion={[style=thick,size=3pt]0.7}] (y2);
				(z1) -- [HQET] (z2);
				(y2) -- [fermion] (x);
				(z2) -- [HQET] (x);
				};
			\end{feynman}
		\end{tikzpicture}+\text{UVCT}}\\&
		\begin{tikzpicture}[baseline=($(p1)!0.5!(x)$)]
			\begin{feynman}
				\vertex (p1);
				\vertex[right=1.2cm of p1] (p2);
				\vertex at ($(p1)!0.3!(p2)+(0,1.8cm)$) (x) {\(\vb{r}\)};
				\vertex at ($(p1)!0.7!(p2)+(0,1.8cm)$) (0) {\(\vb{0}\)};
				\node[square dot] at ($(p1)!0.2!(x)$) (y1);
				\vertex at ($(p2)!0.2!(0)$) (z1);
				\vertex at ($(p1)!0.7!(x)$) (y2);
				\vertex at ($(p2)!0.7!(0)$) (z2);
				\diagram* {
				(y1) -- [scalar] (z1);
				(y2) -- [scalar] (z2);
				(p1) -- [fermion] (y1);
				(p2) -- [HQET] (z1);
				(y1) -- [fermion] (y2);
				(z1) -- [HQET] (z2);
				(y2) -- [fermion] (x);
				(z2) -- [HQET] (0);
				};
			\end{feynman}
		\end{tikzpicture}=
		\pqty{\begin{tikzpicture}[baseline=($(p1)!0.5!(x2)$)]
			\begin{feynman}
				\vertex                  (p1) ;
				\vertex[right=1.2cm of p1] (p2) ;
				\node[crossed dot,minimum size=\dotsize] at ($(p1)!0.5!(p2)+(0,0.4cm)$) (x) ;
				\vertex[above=0.4cm of p1] (p12);
				\vertex[right=1.2cm of p12] (p22);
				\vertex at ($(p12)!0.3!(p22)+(0,1.4cm)$) (x2) {\(\vb{r}\)};
				\vertex at ($(p12)!0.7!(p22)+(0,1.4cm)$) (02) {\(\vb{0}\)};
				\node[square dot] at ($(p12)!0.2!(x2)$) (y12);
				\vertex at ($(p22)!0.2!(02)$) (z12);
				\vertex at ($(p12)!0.6!(x2)$) (y22);
				\vertex at ($(p22)!0.6!(02)$) (z22);
				\diagram* {
				(p1) -- [fermion] (x);
				(p2) -- [HQET] (x);
				};
				\diagram* {
				(y12) -- [very thick, scalar] (z12);
				(y22) -- [very thick, scalar] (z22);
				(y12) -- [very thick,fermion] (y22);
				(z12) -- [very thick, HQET] (z22);
				(y22) -- [very thick,fermion] (x2);
				(z22) -- [very thick, HQET] (02);
				};
			\end{feynman}
		\end{tikzpicture}-\begin{tikzpicture}[baseline=($(p1)!0.5!(x)$)]
			\begin{feynman}
				\vertex (p1);
				\vertex[right=1.2cm of p1] (p2);
				\node[crossed dot,minimum size=\dotsize] at ($(p1)!0.5!(p2)+(0,1.8cm)$) (x) ;
				\node[square dot] at ($(p1)!0.2!(x)$) (y1);
				\vertex at ($(p2)!0.2!(x)$) (z1);
				\vertex at ($(p1)!0.6!(x)$) (y2);
				\vertex at ($(p2)!0.6!(x)$) (z2);
				\vertex at ($(y1)!0.5!(z1)$) (t);
				\diagram* {
				(y1) -- [very thick,scalar] (z1);
				(y2) -- [very thick,scalar] (z2);
				(p1) -- [fermion] (y1);
				(p2) -- [HQET] (z1);
				(y1) -- [very thick,fermion] (y2);
				(z1) -- [very thick,HQET] (z2);
				(y2) -- [very thick,fermion] (x);
				(z2) -- [very thick,HQET] (x);
				};
			\end{feynman}
		\end{tikzpicture}-\text{UVCT}}
		+\pqty{\begin{tikzpicture}[baseline=($(p1)!0.5!(x2)$)]
			\begin{feynman}
				\vertex                  (p1) ;
				\vertex[right=1.2cm of p1] (p2) ;
				\node[crossed dot,minimum size=\dotsize] at ($(p1)!0.5!(p2)+(0,1cm)$) (x) ;
				\node[square dot] at ($(p1)!0.3!(x)$) (y1);
				\vertex at ($(p2)!0.3!(x)$) (z1);
				\vertex[above=1cm of p1] (p12);
				\vertex[right=1.2cm of p12] (p22);
				\vertex at ($(p12)!0.3!(p22)+(0,0.8cm)$) (x2) {\(\vb{r}\)};
				\vertex at ($(p12)!0.7!(p22)+(0,0.8cm)$) (02) {\(\vb{0}\)};
				\vertex at ($(p12)!0.4!(x2)$) (y12);
				\vertex at ($(p22)!0.4!(02)$) (z12);
				\diagram* {
					(y1) -- [scalar] (z1);
					(p1) -- [fermion] (y1);
					(p2) -- [HQET] (z1);
					(y1) -- [fermion] (x);
					(z1) -- [HQET] (x);
				};
				\diagram* {
				(y12) -- [very thick, scalar] (z12);
				(y12) -- [very thick,fermion] (x2);
				(z12) -- [very thick, HQET] (02);
				};
			\end{feynman}
		\end{tikzpicture}-\begin{tikzpicture}[baseline=($(p1)!0.5!(x)$)]
			\begin{feynman}
				\vertex (p1);
				\vertex[right=1.2cm of p1] (p2);
				\node[crossed dot,minimum size=\dotsize] at ($(p1)!0.5!(p2)+(0,1.8cm)$) (x) ;
				\node[square dot] at ($(p1)!0.2!(x)$) (y1);
				\vertex at ($(p2)!0.2!(x)$) (z1);
				\vertex at ($(p1)!0.6!(x)$) (y2);
				\vertex at ($(p2)!0.6!(x)$) (z2);
				\vertex at ($(y1)!0.5!(z1)$) (t);
				\diagram* {
				(y1) -- [scalar] (z1);
				(y2) -- [very thick,scalar] (z2);
				(p1) -- [fermion] (y1);
				(p2) -- [HQET] (z1);
				(y1) -- [fermion] (y2);
				(z1) -- [HQET] (z2);
				(y2) -- [very thick,fermion] (x);
				(z2) -- [very thick,HQET] (x);
				};
			\end{feynman}
		\end{tikzpicture}}+
		\pqty{\begin{tikzpicture}[baseline=($(p1)!0.5!(x)$)]
			\begin{feynman}
				\vertex (p1);
				\vertex[right=1.2cm of p1] (p2);
				\node[crossed dot,minimum size=\dotsize] at ($(p1)!0.5!(p2)+(0,1.8cm)$) (x) ;
				\node[square dot] at ($(p1)!0.2!(x)$) (y1);
				\vertex at ($(p2)!0.2!(x)$) (z1);
				\vertex at ($(p1)!0.6!(x)$) (y2);
				\vertex at ($(p2)!0.6!(x)$) (z2);
				\vertex at ($(y1)!0.5!(z1)$) (t);
				\diagram* {
				(y1) -- [scalar] (z1);
				(y2) -- [scalar] (z2);
				(p1) -- [fermion] (y1);
				(p2) -- [HQET] (z1);
				(y1) -- [fermion] (y2);
				(z1) -- [HQET] (z2);
				(y2) -- [fermion] (x);
				(z2) -- [HQET] (x);
				};
			\end{feynman}
		\end{tikzpicture}+\text{UVCT}}
		\end{split}\end{displaymath}
		\caption{Illustration of the OPE structure of the four-point Green functions through order
	$Z^2\alpha^2$. The first line is for the Wilson coefficient ${\cal C}^{(1)}(r)$,
	the two bottom lines for the Wilson coefficient ${\cal C}^{(2)}(r)$.
	The thick line indicates the corresponding loop momentum to be ``hard'' $(\sim {\bf q})$.}
	\label{Wilson:coeff:coor:space:decomp}
	\end{figure}
	\twocolumngrid

\paragraph{\color{blue} Sketch of derivation of OPE relation \eqref{OPE:NREFT:rel:corr}.}
To deduce the aforementioned OPE relation, it is convenient to consider the
four-point Green function formed by the vacuum matrix element of $\psi({\bf r})N(0)$
multiplied with the Fourier-transformed fields $\widetilde{\psi}^\dagger (p)$ and $\widetilde{N}^\dagger(k)$.
Assuming $p, k \sim mZ\alpha\ll 1/r\sim m $ are soft momenta,
our task is to verify this Green function has the desired factorization property demanded by
\eqref{OPE:NREFT:rel:corr}, order by order in $Z\alpha$.
In \cite{Huang:2018ils}, we have presented a detailed
derivation of the OPE relation through order $Z^2\alpha^2$ for Klein-Gordon case,
with the technicalities expounded.
The calculation in \cite{Huang:2018ils} can be directly carried over here,
supplemented merely with the extra complication of including the Darwin term.

In Fig.~\ref{Wilson:coeff:coor:space:decomp}, we pictorially show the procedure to deduce the
intended coefficient ${\cal C}(r)$ through order $Z^2\alpha^2$, by subtracting and adding back the
Green function inserted with the local operator $\psi N$, and expanding the integrand
in the hard sub-loop region  (our method is analogous to the Wilson expansion
for the product of two $\phi$ fields in the $\lambda \phi^4$ theory~\cite{Collins:1984xc}).
Here we just present the final results in MS scheme:
\begin{align}
& \mathcal{C}_{\ref{Wilson:coeff:coor:space:decomp}b)}= -Z^2\alpha ^2
\Bqty{\ln \mu r+\frac{1}{2} \ln {4\pi e^{\gamma_E -1}}},
\label{Wilson:coeff:Cr:kin:Dar}\\
& \mathcal{C}_{\ref{Wilson:coeff:coor:space:decomp}c)}= {Z^2\alpha ^2\over 2}
\left\{\ln \mu r + {1\over 4}\left[\gamma_E + 2\ln 2\pi - \psi\left({3\over 2}\right) \right]\right\},
\nn
\end{align}
where $\gamma_E$ is the Euler constant, $\psi$ is the di-$\Gamma$ function.
The first term has already been given in \cite{Huang:2018ils}, the second is new.
Summing up these two terms, we then obtain the Wilson coefficient
$-{Z^2\alpha^2\over 2}\ln \mu r$ recorded in \eqref{OPE:NREFT:rel:corr}.

\paragraph{\color{blue} RGE for Wilson coefficient ${\cal C}(r)$.}
It seems advantageous to invoke the powerful renormalization group equation (RGE) to resum large
logarithms appearing in the Wilson coefficient in \eqref{OPE:NREFT:rel:corr} to all orders in $Z\alpha$,
effectively explain the anomalous scaling in \eqref{Fn:asym:small:r} for Dirac wave function.
Demanding the left hand side of \eqref{OPE:NREFT:rel:corr} is independent of $\mu$,
we then write down the RGE for ${\cal C}(r)$:
\begin{align}
\mu {\partial {\cal C}(r,\mu)\over \partial \mu}+  \gamma_{\mathcal S} {\cal C}(r,\mu) = 0,
\label{RGE:cal:C:r:mu}
\end{align}
with the anomalous dimension given in \eqref{ano-dim}.
Since $\mu$ and $r$ are glued together inside the logarithms,
dimensional analysis leads to $(\mu {\partial \over \partial\mu} - r {\partial \over \partial r}){\cal C}=0$.
We can transform the RGE \eqref{RGE:cal:C:r:mu} into
the evolution equation that controls the scaling of ${\cal C}$ in small $r$ reads
\begin{align}
r {\partial {\cal C} \over \partial r}+ \gamma_{\mathcal S} {\cal C} = 0.
\label{RGE:cal:C:r:kappa}
\end{align}

The solution of \eqref{RGE:cal:C:r:kappa} becomes
\begin{align}
{\cal C}(r,\mu) = {\cal C}(r_0,\mu) \left({r\over r_0}\right)^{-{Z^2\alpha^2\over 2}}.
\label{RGE:cal:C:r}
\end{align}
As anticipated, after resumming the leading logarithms,
the RGE-improved Wilson coefficient ${\cal C}(r)$
fully agrees with the near-the-origin behavior of Dirac wave function in
\eqref{Fn:asym:small:r}.

It is enlightening to pursue the closer contact between our field-theoretical machinery
and \eqref{Fn:asym:small:r}.
Let us define a reference length as $r_0={n a_0\over 2}$ for the $nS$ hydrogen state, and $\mu_0=1/r_0$.
The boundary condition is ${\cal C}(r=r_0;\mu=\mu_0)=1$.
In accord with \eqref{Dirac:wf:nonlocal:matrix:element:NREFT:field}),
provided that we identify the following local operator
matrix element with the nonrelativistic Schrodinger-Pauli wave function at the origin:
\begin{align}
\langle 0\vert [\psi N]_R(0;\mu_0)\vert n S_{1/2},m  \rangle \approx  {1\over \sqrt{4\pi}}\,R_{n0}^{\rm Sch}(0)\,\xi_m,
\end{align}
we are then able to fully recover \eqref{Fn:asym:small:r}.

\paragraph{\color{blue}Summary.}
The Coulomb solution to Dirac equation marks an milestone in the history of quantum mechanics.
Until very recently, the physics underlying the divergent near-the-origin behavior of
Dirac wave function for $nS_{1/2}$ hydrogen is rather obscure.

Together with our recent work~\cite{Huang:2018ils}, we employ the modern field-theoretical machinery such as
EFT and OPE to solve the long-standing conundrums related to hydrogen KG and Dirac wave functions.
The key is to construct the nonrelativistic EFT for atoms by combining NRQED and HQET.
By incorporating the relativistic kinetic correction and Darwin term, we show that
the universal logarithmic near-the-origin divergence can be simply interpreted as the Wilson coefficients
arising from the product of the electron and nucleus fields.  RGE further allows us to
recover the anomalous scaling behavior observed in the KG and Dirac wave functions at small $r$.
By our approach, we can readily trace the different scaling behavior of these two types of
relativistic wave functions is due to Darwin term.

A virtue of our field-theoretical approach is that, since OPE holds at operator level,
our main result \eqref{OPE:NREFT:rel:corr} can apply to any multi-electron atom as well,
not necessarily confined for hydrogen-like atoms.

The folklore says that~\cite{Itzykson:1980rh}, as the solutions to the relativistic wave equations,
the KG and Dirac wave functions in Coulomb field can be probed in arbitrarily small distance.
Our OPE analysis based on relativistic QED utterly fails to reproduce the true short-distance
divergence of the KG and Dirac wave functions.
In our opinion, this symptom may indicate that the KG and Dirac equations of hydrogen
are actually secretely {\it nonrelativistic} bound-state equations.
It does not make sense to talk about the near-the-origin behavior of the KG and Dirac
wave functions when $r\ll {1\over m}$.  When probing the short-range behavior of these
wave functions, the minimum value of $r$ should be frozen around ${1\over m}$,
since the electron Compton length of constitutes the smallest length scale
where nonrelativistic EFT may still apply.


\begin{acknowledgments}
{\noindent\it Acknowledgment.}
One of the authors (Y.~J.) thanks Xiangdong Ji for spurring his interest in investigating
the near-the-origin divergence of hydrogen Dirac wave function during his visit at University of Maryland in 2011,
and especially for the emphasis that this phenomenon should be linked with renormalization.
This work is supported in part by the National Natural Science Foundation of China under Grants No.~11875263,
No.~11621131001 (CRC110 by DFG and NSFC).
\end{acknowledgments}

\appendix

\end{document}